\documentclass[a4paper]{article}

\usepackage{INTERSPEECH2022}

\title{Speaker Characterization by means of Attention Pooling}
\name{Federico Costa, Miquel India, Javier Hernando}
\address{
  TALP Research Center, Department of Signal Theory and Communications,\\
  Universitat Politecnica de Catalunya, Barcelona, Spain}
\email{federico.costa@upc.edu, miquel.angel.india@upc.edu, javier.hernando@upc.edu}

\begin{document}

\maketitle
\begin{abstract}

State-of-the-art Deep Learning systems for speaker verification are commonly based on speaker embedding extractors. 
These architectures are usually composed of a feature extractor front-end together with a pooling layer to encode variable-length utterances into fixed-length speaker vectors. 
The authors have recently proposed the use of a Double Multi-Head Self-Attention pooling for speaker recognition, placed between a CNN-based front-end and a set of fully connected layers. 
This has shown to be an excellent approach to efficiently select the most relevant features captured by the front-end from the speech signal.
In this paper we show excellent experimental results by adapting this architecture to other different speaker characterization tasks, such as emotion recognition, sex classification and COVID-19 detection.

\end{abstract}
\noindent\textbf{Index Terms}: multi-head self-attention, double attention, speech recognition, speaker verification, speaker characterization

\section{Introduction}

Speaker Verification (SV) aims to determine whether a pair of audios corresponds to the same speaker. 
Given speech signals, speaker verification systems are able to extract speaker identity patterns from the characteristics of the voice. 
State-of-the-art SV systems are commonly Deep Learning (DL) approaches that encode speaker characteristics into discriminative speaker vectors (also known as speaker embeddings).
These architectures are usually trained as speaker classifiers in order to be used as speaker embedding extractors.
One of the most known speaker representations is the x-vector \cite{Snyder18-XVR}, which has become state-of-the-art for speaker recognition.
Recent network architectures used for speaker embedding generation are composed of a front-end feature extractor, a pooling layer and a set of Fully Connected (FC) layers.
During the last years, several studies addressed different types of pooling strategies \cite{Cai18-ETE, Xie19-ULA, Jung19-SPE}. 
X-vector originally uses statistical pooling \cite{Snyder17-DNN}. 
Self-attention mechanisms have been used to improve statistical pooling, such as \cite{Zhu18-SAS, India19-SMH}.

Some of these speaker embedding generation systems can be adapted and used to solve Speaker Characterization (SC) tasks \cite{Snyder18-SLR, Pappagari20-XVM}.
SC tasks are those where one or more speaker characteristics are extracted from the speech.
In the last years, DL approaches were used in order to extract speaker characteristics from speech such as emotions \cite{Meng19-SER}, sex \cite{Alnuaim22-SGR}, age \cite{Tursunov15-AAG}, language \cite{Bartz17-LIU}, dialect \cite{Wang21-AET}, accent \cite{Weninger19-DLB}, health conditions \cite{Cummins18-SAF}, among others.

The authors have recently proposed a DL system based on a Double Multi-Head Self-Attention (DMHSA) pooling \cite{India20-DMH}.
Its architecture consists of a Convolutional Neural Network (CNN)-based front-end, followed by an attention-based pooling layer and a set of fully connected layers.
It has been proven to achieve excellent results in SV tasks and it is an improvement of the Multi-Head Self-Attention (MHSA) pooling method proposed in \cite{India19-SMH}.
Since this system is trained as a speaker classifier, it can be adapted to solve SC tasks.
In this paper we will present several applications using an adapted version of this architecture on different SC tasks, namely Speech Emotion Recognition (SER), Speaker Sex Classification and Speaker COVID-19 Detection, achieving excellent results.
We would like to mention that all Speaker COVID-19 Detection results obtained in this work are meant for exploratory purposes and should not be taken to draw any medical conclusions.

The rest of this paper is structured as follows.
Section 2 explains DMHSA pooling. 
Section 3 gives the details of the DMHSA system applied to SV, with the experimental setup and results included.
Section 4 gives the details of the DMHSA system applied to SC, with several applications and their experimental setup and results included.
The concluding remarks are given in Section 5.

\section{Double Multi-Head Self-Attention Pooling}

\begin{figure}[h!]
  \centering
  \includegraphics[width=\linewidth]{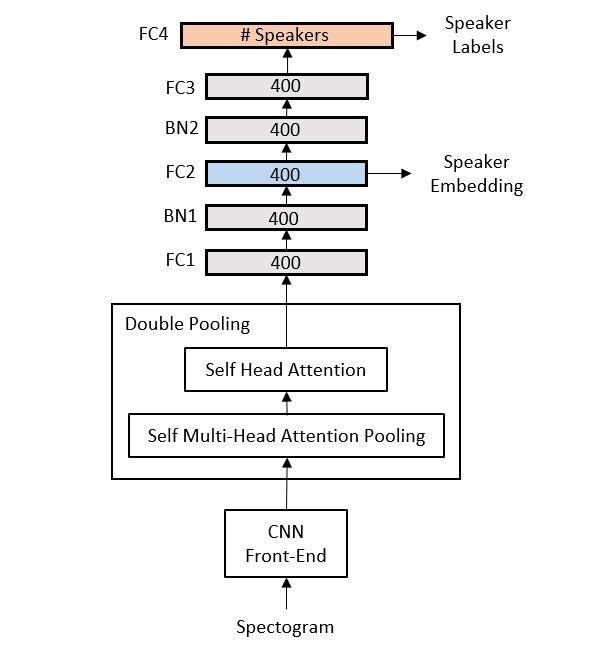}
  \caption{DMHA architecture.}
  \label{fig:dmha_architecture}
\end{figure}

In \cite{India20-DMH}, DMHSA Pooling was proposed. It is a DL system with an attention-based pooling layer that was developed for SV. 
This model is trained as a speaker classifier, with the capability of learning effective discriminative speaker embeddings.
The system architecture is illustrated in Figure~\ref{fig:dmha_architecture}. 

The proposed network was trained to classify variable-length speaker utterances. 
The system uses as input a spectrogram of the input audio, using the log Mel Spectrogram, with 25 milliseconds length Hamming windows and 10 milliseconds window shift.  
It accepts variable length input of size $N$ (in frames) and has a fixed number of Mel bands $M=80$. 
Then, the input resulting data is of size $N\text{x}80$.
The audio features have been normalized with Cepstral Mean Normalization. 

\subsection{CNN front-end feature extractor}

The feature extractor network is an adapted version of the VGG proposed in \cite{Hori17-AIJ}.
It is fed with $N\text{x}80$ spectrograms to obtain a sequence of encoded hidden representations
This CNN comprises 4 convolutional blocks, each of which contains two contatenated convolutional layers followed by a max pooling layer. 
Each convolutional layer has a set of $3$x$3$ filters with a stride of 1 (a same convolution). 
Each max pooling layer consists of a $2$x$2$ max pooling with a stride of 2 (and no padding). 
The first convolutional block applies a set of 128 filters, the second applies 256 filters, the third applies 512 filters and the fourth applies 1024 filters.
Hence, given a spectrogram of $N$ frames and $M$ features, the VGG performs a down-sampling reducing its output into a tensor of $\frac{N}{16}$x$\frac{M}{16}$x$D'$ dimension, where $D'$ is the final quantity of channels. 
This tensor is reshaped into a sequence of hidden states $h_i \in \mathbb{R}^{D}$, with $i=1, ..., \frac{N}{16}$ and $D = \frac{M}{16}\text{x}D'$.

\subsection{Double Multi-Head Self Attention pooling}

From the feature extractor, we get a sequence of hidden states $h_1, ..., h_{N/16} \in \mathbb{R}^{D}$.
If we consider a number of $K$ heads for the MHSA pooling, we split each hidden state $h_t$ into $K$ new hidden states so that $h_t = [h_{t1}, ..., h_{tK}]$, where $h_{tj} \in \mathbb{R}^{D/K}$.
Now, for each head $j$, a self-attention operation is applied over the encoded sequences $[h_{1j}, ..., h_{\frac{N}{16}j}]$.
The weights of each head alignment are defined as:
\begin{equation}
  w_{tj} = \frac{\text{exp} \left( \frac{h_{tj}^T u_j}{\sqrt{d_h}} \right)}{\sum_{l=1}^{N/16}  \text{exp} \left( \frac{h_{lj}^T u_j}{\sqrt{d_h}} \right)}
  \label{eq1}
\end{equation}
where \( w_{tj} \) corresponds to the attention weight of the head \( j \) on the step \( t \) of the sequence, \( u_j \in \mathbb{R}^{D/K} \) is a trainable parameter and \( d_h \) corresponds to the hidden state dimension \( D/K \).
We then compute a new pooled representation for each head:
\begin{equation}
  c_{j} = \sum_{t=1}^{N/16} h_{tj}^T w_{tj}
  \label{eq2}
\end{equation}
where \( c_{j} \in \mathbb{R}^{D/K} \) corresponds to the utterance level representation from head j.
Each self-attention operation for the head $j$ can be understood as a dot-product attention where the keys and the values correspond to the same representation $[h_{1j}, ..., h_{\frac{N}{16}j}]$ and the query is only a trainable parameter $u_j$.

Self-attention is now used to pool the set of head context vectors $c_i$ in order to obtain an overall context vector $c$:
\begin{equation}
  w_i^{'} = \frac{\text{exp} \left( c_{i}^T u^{'} \right)}{\sum_{l=1}^{K} \text{exp} \left( c_{l}^T u^{'} \right)}
  \label{eq3}
\end{equation}
\begin{equation}
  c = \sum_{t=1}^{K} c_{i}^T w_i^{'}
  \label{eq4}
\end{equation}
where \( w_{i}^{'} \) corresponds to the aligned weight of each head and \( u^{'} \in \mathbb{R}^{D/K} \) is a trainable parameter.

With this method, each utterance context vector $c$ is computed as a weighted average of the context vectors among heads.  
This allows to capture different kinds of speaker patterns in different regions of the input and, at the same time, to weigh the relevance of each of these patterns for each utterance.

\subsection{Fully-connected layers}
The utterance-level speaker vector obtained from the pooling layer is fed into a set of four FC layers.  
Each of the first two FC layers is followed by a batch normalization and Rectified Linear Unit (ReLU) activations. 
A dense layer is adopted for the third FC layer and the last FC is a SoftMax layer that corresponds to the speaker classification output layer.

Once the network is trained, we can extract a speaker embedding from one of the intermediate FC layers. 
According to \cite{Liu19-LMS}, we consider the second layer as the speaker embedding instead of the third one. 
The output of this layer then corresponds to the speaker representation that will be used for the speaker verification task.

\section{Speaker Verification}
\subsection{Experimental setup}

The DMHSA system has been assessed in \cite{India20-DMH} by VoxCeleb dataset \cite{Chung18-VDS, Nagrani18-VAL}. 
VoxCeleb is a large multimedia database that contains more than one million 16kHz audio utterances for more than 6K celebrities and has two different versions with several evaluation protocols. 
For \cite{India20-DMH} experiments, VoxCeleb2 development partition with no augmentation has been used to train all models. 
The performance of these systems has been evaluated with Vox1-Test, Vox1-E, and Vox1-H conditions. 
Vox1 test only uses the test set, Vox1-E uses the whole development + test corpus and Vox1-H is restricted to audio pairs from same nationality and gender speakers.

DMHSA pooling has been evaluated against two self-attentive-based pooling methods: MHSA \cite{India19-SMH} and vanilla Self-Attention (which is indeed a single-head MHSA).
In order to evaluate them, only the pooling layer of the system (Figure~\ref{fig:dmha_architecture}) has been replaced without modifying any other block or parameter from the network. 
The speaker embeddings used for the verification tests have been extracted from the same FC layer for each of the pooling methods.
Cosine distance has been used to compute the scores between pairs of speaker embeddings.
The number of heads for both MHSA and DMHSA pooling layer were tuned: 8, 16, and 32 heads were considered.

\subsection{Results}

The results of \cite{India20-DMH} experiments are shown in Table~\ref{tab:dmha_results}.
Performance was evaluated using Equal Error Rate (EER).
Both Self-Attention and MHSA approaches were used as baselines, since they have been proved to outperform three previous baselines: statistical and temporal pooling based methods and an i-vector + PLDA system \cite{India19-SMH}.
Self-Attention pooling has shown very similar results compared to MHSA approaches.
DMHSA have shown better results for all head values compared with both Self Attention and MHSA approaches.
Best performance in DMHSA based models has been achieved with 16 and 32 heads. 
Within Vox1-Test protocol, the best DMHSA model (16 heads) has shown a $6.73\%$ and $5.06\%$ relative improvement in terms of EER compared to Self Attention pooling and the best MHSA pooling (8 heads), respectively.

\begin{table*}[h]
  \caption{Speaker verification evaluation results on VoxCeleb 1 protocols.}
  \label{tab:dmha_results}
  \centering
  \begin{tabular}{ r@{} l c c c c }
    \toprule
    \multicolumn{2}{c}{\textbf{Approach}} & \multicolumn{1}{c}{\textbf{Heads}} & \multicolumn{1}{c}{\textbf{Vox1-Test EER}} & \multicolumn{1}{c}{\textbf{Vox1-E EER}} & \multicolumn{1}{c}{\textbf{Vox1-H EER}} \\
    \midrule
    & Self-Attention    &   1   &   $3.42$   &   $3.42$   & $4.89$  \\
    & MHSA              &   8   &   $3.36$   &   $3.44$   & $5.04$  \\
    & MHSA              &   16  &   $3.43$   &   $3.40$   & $4.90$  \\
    & MHSA              &   32  &   $3.64$   &   $3.68$   & $5.35$  \\
    & DMHSA             &   8   &   $3.27$   &   $3.23$   & $4.69$  \\
    & DMHSA             &   16  &   $3.19$   &   $3.22$   & $4.67$  \\
    & DMHSA             &   32  &   $3.23$   &   $3.18$   & $4.61$  \\
    \bottomrule
  \end{tabular}
  
\end{table*}

\section{Speaker characterization applications}

In this section we present several works that adapted and applied the DMHSA system to SC tasks, which are: \cite{Aromi21-PEI} for SER, \cite{Garriga22-DLF} for Speaker Sex Classification and \cite{Marchan22-DEI} for Speaker COVID-19 Detection.

\subsection{Speaker Emotion Recognition}

SER is an SC task where the goal is to recognize the emotion from speech.
Human emotion is classified using the six archetypal emotions approach, from which all emotional states can be derived \cite{Cowie03-DTE}. 
These archetypal emotions are Anger, Fear, Disgust, Surprise, Joy, Sadness and an added Neutral state.
Several attempts at SER using statistical models have provided acceptable results, such as Gaussian Mixture Models (GMM) \cite{ElAyadi07-SER}, Hidden Markov Models (HMM) \cite{Nogueiras01-SER} and i-vectors \cite{Xia12-UIV}. 
Over the last years, SER has really taken off with the explosion of DL techniques \cite{Kim17-TSE, Meng19-SER, Li19-IET}.

\subsubsection{Experimental setup}

In \cite{Aromi21-PEI}, the DMHSA system was adapted to solve SER.
It was assessed by the INTERFACE dataset \cite{Hozjan02-IDD}. 
This database was designed for the general study of emotional speech. 
It contains recordings in four different languages: Slovenian, English, Spanish and French, distributed uniformly. 
For each language, there are 170-190 sentences spoken by two different actors, one male and one female, except for Spanish where two male actors and one female are used. 
Each sentence is spoken in seven different styles, for a total number of 24,197 recordings. 
The styles (emotion labels) are: Anger, Sadness, Joy, Fear, Disgust, Surprise and a Neutral style with different variations depending on the language. 
For the purpose of classification, all these styles were considered as one general “Neutral” style. 
Therefore, $28\%$ of the samples were labelled as Neutral style. 
The rest of the samples are uniformly distributed between the other labels.

The DMHSA model was adapted to this problem by changing the SoftMax layer to a seven units output (one for each emotion label) and by using a 3-blocks front-end feature extractor instead of the 4-blocks one proposed in \cite{India20-DMH}.
Several models with different attention mechanisms (vanilla Self-Attention, MHSA and DMHSA) were trained: 1 for Self-Attention, 5 for MHSA and 5 for DMHSA, both with different head numbers (4, 8, 16, 32 and 64).
These models were tested against a baseline model, which architecture had a statistical pooling component instead of an attention-based one.

To train the model, the entirety of the dataset was split into three parts: $70\%$ train, $10\%$ validation and $20\%$ test.
The training was done with early stopping parameter equal to 5, batch size was 64 and the number of seconds of temporal window taken from the audio files in training was fixed to 1 second. 
For validation and test the whole utterances have been encoded. 
Adam optimizer was used, the learning rate was set to $0.0001$ and the weight decay parameter was set to $0.01$.
In the case of DMHSA, a head drop probability of 0.3 was fixed for 16, 32 and 64 heads, meaning that any given head had a probability of $30\%$ to have its weight set to 0. 
The drop probability for 4 and 8 heads was set to $0.01$ because the number of heads was too low for a $30\%$ drop and caused instabilities in training.

\subsubsection{Results}

The results of \cite{Aromi21-PEI} experiments are shown in Table~\ref{tab:aromi_ser_results}.
Performance was evaluated on the test set using Accuracy. 
Self-attention performed close to statistical pooling, but slightly worse.
For MHSA, the best results were obtained with 32 heads, with a relative accuracy improvement of $1.23\%$ with respect to statistical pooling. 
For DMHSA, the best result was obtained again with 32 heads, but slightly worse than statistical pooling accuracy. 
MHSA has effectively obtained a better pooled representation than statistical pooling by being more selective about the information contained in the embedding.
The reason for the underwhelming performance of Self-Attention could be the short 1 second temporal window used, since if the input contains more frames, more attention weights can be assigned and this could lead to a better pooled representation.
While DMHSA provides an increase in performance for SV, the compression factor that comes with the DMHSA seems to affect the model’s ability to predict emotion in this experiment (compared to MHSA which does not compress the final context vector).

\begin{table}[th]
  \caption{SER test results from \cite{Aromi21-PEI}}
  \label{tab:aromi_ser_results}
  \centering
  \begin{tabular}{ r@{} l  c } 
    \toprule
    \multicolumn{2}{c}{\textbf{Model}} & \multicolumn{1}{c}{\textbf{Accuracy}} \\
    \midrule
    &   MHSA 32 Heads       &   $91.09\%$   \\
    &   Statistical Pooling &   $89.98\%$   \\
    &   DMHSA 32 Heads      &   $89.87\%$   \\
    &   Self-Attention      &   $89.32\%$   \\
    \bottomrule
  \end{tabular}
  
\end{table}

\subsection{Speaker Sex Classification}

Speaker Sex Classification is an SC task where the speaker's sex is extracted from a voice utterance.
Over the last years, multiple investigations have tried to fully optimize Speaker Sex Classification systems' performance using DL in order to accomplish almost perfect results \cite{Buyukyilmaz16-VGR, Ertam19-AEG, Tursunov15-AAG, Alnuaim22-SGR}.

\subsubsection{Experimental setup}

In \cite{Garriga22-DLF}, the DMHSA system was adapted to solve a speaker sex classification task.
It was assessed by the Catalan Common Voice dataset \cite{CVM22-CCV}. 
The Catalan Common Voice dataset was created in 2018 and has seen a huge expanding thanks to the promotion it has received in the last years and the voice donations of the Catalan-speaking population.
The creation of this database and the need to adapt artificial intelligence technologies to the Catalan language, lead to the development of the AINA project \cite{Boersma20-ALN}. 
This project -promoted by the Departament de la Vicepresidència i de Polítiques Digitals i Territori of the Catalan government and the Barcelona Supercomputing Center- aims to shape speech recognition techniques to Catalan.
For sex classification, three labels are given: Male (with 385,061 samples), Female (with 117,666 samples) and Other (with 418 samples). 
A significant imbalance towards the Male class can be observed. 
The class under the label Other was discarded.
For this task’s experiment, as the number of samples from each class in the dataset was large enough, almost completely balanced classes were used for the training set (around 70,000 samples from each class).
Only utterances longer than 2.5 seconds were considered.

The DMHSA model was adapted to this problem by changing the SoftMax layer to a two units output (one for each sex label) and by using a 3-blocks front-end feature extractor.
The DMHSA model was trained and tested with 32 heads and two different loss functions: Cross-Entropy (CE) and Weighted Cross-Entropy (WCE).
To train the model, the entirety of the dataset was split into train, validation and test.
The training was done with early stopping parameter equal to 5, batch size was 64 and the number of seconds of temporal window taken from the audio files in training was fixed to 2.5 seconds. 
Adam optimizer was used, the learning rate was set to $0.0001$ and the weight decay parameter was set to $0.001$.

\subsubsection{Results}

The results of \cite{Garriga22-DLF} experiments are shown in Table~\ref{tab:garriga_sex_results}.
Performance was evaluated on the test set using Accuracy and F-score.
Both DMHSA with CE and WCE performed with excellent results, above $95\%$ accuracy and with a 0.96 f-score.
As the classes were already almost perfectly balanced, using the WCE loss did not represent any significant improvements.

\begin{table}[th]
  \caption{Speaker sex classification test results from \cite{Garriga22-DLF}}
  \label{tab:garriga_sex_results}
  \centering
  \begin{tabular}{ r@{} l  c  c}
    \toprule
    \multicolumn{2}{c}{\textbf{Model}} &  \multicolumn{1}{c}{\textbf{Accuracy}} &  \multicolumn{1}{c}{\textbf{F-Score}} \\
    \midrule
    &   DMHSA 32 heads + CE Loss    &   $95.60\%$   &   $0.96$  \\
    &   DMHSA 32 heads + WCE Loss   &   $95.80\%$   &   $0.96$  \\
    \bottomrule
  \end{tabular}
  
\end{table}

\subsection{Speaker COVID-19 Detection}

Speaker COVID-19 Detection is the SC task of determining the presence of a COVID-19 infection by analyzing the generated sounds from the respiratory system (whether cough, breathing or regular speech).
Recently, several COVID-19 detection from speech DL approaches have been developed \cite{Nessiem21-DC1, Nassif22-C1D}.
We would like to remind that all Speaker COVID-19 Detection results obtained in this work are meant for exploratory purposes and should not be taken to draw any medical conclusions.

\subsubsection{Experimental setup}

In \cite{Marchan22-DEI}, the DMHSA system was adapted to solve a COVID-19 detection task using cough audios.
It was assessed by the CCS dataset, which is part of the database that was used in the INTERSPEECH 2021 Computational Paralinguistics ChallengE (ComParE) \cite{Schuller21-TI2}.
This database was already split into train, validation and test sets. 
There are in total 725 audios, which 158 are samples from COVID-19 positive diagnostic and 567 are negative samples.
There are 286 audios in the train set, 231 audios in the validation set and 208 audios in the test set.
The test set was cleaned and filtered in order to improve the quality of the samples.

The DMHSA model was adapted to this problem by changing the SoftMax layer to a two units output (one for each diagnostic label) and by using a 3-blocks front-end feature extractor.
Two 32-heads DMHSA models were trained using two different loss functions: CE and WCE.
The training was done with early stopping parameter equal to 50 and a batch size of 64.
Adam optimizer was used, the learning rate was set to $0.0001$ and the weight decay parameter was set to $0.001$.
A head drop probability of 0.3 was fixed for the heads, meaning that any given head had a probability of $30\%$ to have its weight set to 0. 

\subsubsection{Results}

The results of \cite{Marchan22-DEI} experiments are shown in Table~\ref{tab:marchan_covid_results}.
Performance was evaluated on the test set using Area Under the receiver operating characteristic Curve (AUC).
DMHSA with 32 heads and WCE loss performed better than DMHSA with 32 heads and CE loss, with a relative AUC improvement of $8.43\%$.
In this case, DMHSA with 32 heads and WCE achieved very good results, being able to detect effectively COVID-19 samples.
Because of the imbalanced classes, using the WCE loss function improved the results drastically.

\begin{table}[th]
  \caption{COVID-19 Detection test results from \cite{Marchan22-DEI}}
  \label{tab:marchan_covid_results}
  \centering
  \begin{tabular}{ r@{} l  c }
    \toprule
    \multicolumn{2}{c}{\textbf{Model}} & 
                                         \multicolumn{1}{c}{\textbf{AUC}} \\
    \midrule
    &   DMHSA 32 heads + CE Loss    &   $83.00\%$   \\
    &   DMHSA 32 heads + WCE Loss   &   $90.00\%$   \\
    \bottomrule
  \end{tabular}
  
\end{table}

\section{Conclusions}

In this paper we have described a Double Multi-Head Self-Attention pooling mechanism for speaker recognition.
Its architecture consists of a CNN-based front-end, followed by an attention-based pooling layer and a set of fully connected layers.
The network is trained as a speaker classifier and a bottleneck layer from these fully connected layers is used as speaker embedding.
The presented approach has been evaluated in a text-independent speaker verification task using the speaker embeddings and applying the cosine distance.
It has outperformed both vanilla Self Attention and Multi-Head Self-Attention pooling baseline methods.
It also has been adapted to solve several speaker characterization tasks, namely speaker emotion recognition, sex classification and COVID-19 detection, obtaining excellent results.

\section{Acknowledgements}
This work was supported by the project PID2019-107579RBI00, funded by the Spanish Ministry of Science and Innovation.

\newpage

\bibliographystyle{IEEEtran}
\bibliography{paper}

\end{document}